\documentstyle[epsf]{article}          
\catcode`@=11

\@addtoreset{equation}{section}
\catcode`@=12

\begin{document}

\title{{\small\centerline{May 1999 \hfill SINP/TNP/99-22}}
\medskip 
{\bf Determination of cosmological parameters:\\ an introduction for
non-specialists\thanks{Invited talk at the ``Discussion meeting on
Recent Developments in Neutrino Physics'', held at the Physical
Research Laboratory, Ahmedabad, February 2--4, 1999.}}}

\author{\bf Palash B. Pal\\ 
\normalsize Saha Institute of Nuclear Physics, 1/AF Bidhan-Nagar,
Calcutta 700064, India} 

\date{}
\maketitle 

\begin{abstract} \normalsize\noindent
I start by defining the cosmological parameters $H_0$, $\Omega_m$ and
$\Omega_\Lambda$. Then I show how the age of the universe depends on
them, followed by the evolution of the scale parameter of the universe
for various values of the density parameters. Then I define strategies
for measuring them, and show the results for the recent determination
of these parameters from measurements on supernovas of type
1a. Implications for particle physics is briefly discussed at the end.
\end{abstract}
\bigskip\bigskip

\section{Introduction}
In this conference on Neutrino Physics, I have been asked to talk
about the determination of cosmological parameters. The reason for
this, obviously, is the potential importance of neutrinos for
cosmology. They can serve as a component for the dark matter of the
universe. On the other hand, many important constraints on neutrino
properties are derived from various cosmological considerations. The
precise values of such constraints are determined by various
cosmological parameters like the Hubble parameter, the density of
matter in the universe, etc. In the light of this relationship, it is
important to know the cosmological parameters accurately. In this
talk, I will outline the methods for the determination of cosmological
parameters. Since I am not a specialist in this field, at best I can
hope that this exposition will be useful to other non-specialists.

The outline of the paper is the following. In Sec.~\ref{df}, I define
the cosmological parameters. There are, in fact, three of them --- the
Hubble parameter, and the present density of matter energy and vacuum
energy in the universe. In Sec.~\ref{t0}, I show how the age of the
universe depends on these parameters. Then in Sec.~\ref{ev}, I show
how the evolution of the universe depends on the values of the
cosmological parameters. An understanding of the evolution is
essential in formulating strategies for the cosmological
parameters. Some such strategies are discussed in general terms in
Sec.~\ref{st}. At the end of this section, we indicate the outline for
the rest of the paper.

\section{Defining the cosmological parameters}\label{df}
In cosmology, we start with the the assumption of a homogeneous and
isotropic universe, which is described by the
Friedman-Robertson-Walker (FRW) metric:
\begin{eqnarray}
ds^2 = dt^2 - a^2(t) \Bigg[ {dr^2 \over 1-kr^2} + r^2 \Big( d\theta^2
+ \sin^2\theta \, d\phi^2 \Big) \Bigg] \,.
\label{FRWmetric}
\end{eqnarray}
Here, $r$ is a dimensionless co-ordinate distance and $a(t)$ is an
overall scale parameter. The physical distance between two points
depends on the co-ordinate distance and the scale parameter, in a
manner which will be discussed later. The parameter $k$, if it is
non-zero, can always be adjusted to have unit magnitude by adjusting
the definition of $r$. Thus, the possible values of $k$ are given by
\begin{eqnarray}
k = 0, \pm 1 \,.
\end{eqnarray}
Usually, we think of the universe to be open if $k<0$, for which it
expands forever. Contrarily, if $k>0$, the universe is closed, i.e.,
it is destined to recollapse at some time in the future. The
borderline case, $k=0$, is called a flat universe. We will see that
these statements are correct only if there is no cosmological
constant. {}From the general consideration of homogeneity and
isotropy, however, the cosmological constant can be present, and these
notions get modified, as we will see.

In presence of a cosmological constant $\Lambda$, the equations of
motion are given by
\begin{eqnarray}
R_{\mu\nu} - \frac12 g_{\mu\nu} R = 8\pi G T_{\mu\nu} + \Lambda
g_{\mu\nu} \,.
\label{Einstein}
\end{eqnarray}
Here, $R_{\mu\nu}$ is the Ricci tensor which is defined through the
metric of Eq.\ (\ref{FRWmetric}), and $R=R_{\mu\nu}g^{\mu\nu}$.
Although in general Eq.\ (\ref{Einstein}) gives 10 equations, here
most of them are identical because the metric is homogeneous and
isotropic. In fact, one gets only two independent equations. One of
them is given by
\begin{eqnarray}
\left( {\dot a \over a} \right)^2 = {8\pi G\over 3} \, \rho +
{\Lambda\over 3} - {k\over a^2} \,,
\label{evoleqn}
\end{eqnarray}
where $\rho$ is the energy density of matter. This equation is valid
for all time. Applying it to the present time, we can write
\begin{eqnarray}
H_0^2 = {8\pi G\over 3} \, \rho_0 +
{\Lambda\over 3} - {k\over a_0^2} \,,
\label{eqnforH0}
\end{eqnarray}
where $H_0$ is the value of $\dot a/a$ at the present time, and the
subscript `0' on $\rho$ and $a$ denote the values of these parameters
at the present time. We now introduce the dimensionless parameters
\begin{eqnarray}
\Omega_m &\equiv & {8\pi G\over 3H_0^2} \, \rho_0 \,,\nonumber\\* 
\Omega_\Lambda &\equiv & {\Lambda\over 3H_0^2} \,,\nonumber\\* 
\Omega_k &\equiv & -\; {k\over a_0^2H_0^2} \,.
\label{Omegas}
\end{eqnarray}
These definitions enable us to rewrite Eq.\ (\ref{eqnforH0}) in the
following form:
\begin{eqnarray}
1 = \Omega_m + \Omega_\Lambda + \Omega_k \,.
\label{sum=1}
\end{eqnarray}
This shows that, among the three parameters introduced in Eq.\
(\ref{Omegas}), we can take only $\Omega_m$ and $\Omega_\Lambda$ to be
independent. $\Omega_k$ is not. In addition, $H_0$ is another
parameter. These are the three cosmological parameters, and the aim of
this lecture is to show how they may be determined.

Before proceeding, I would like to make one comment. Note that all
these parameters are defined using values of physical quantities {\em
at the present time}. One may tend to think that because of that, the
present time $t_0$ should also be added to the list of parameters. But
we will show that in fact $t_0$ is not independent.

\section{The age of the universe}\label{t0}
So far, we have discussed only one of the independent equations that
arise among the set given in Eq.\ (\ref{Einstein}). In order to
proceed, we need the other. This is in fact equivalent to the
statement of conservation of matter, which means that the quantity
$\rho a^3$ is constant over time, or
\begin{eqnarray}
\rho a^3 = \rho_0 a_0^3 \,.
\end{eqnarray}

From now on, it is more convenient to use dimensionless variables
instead of $a$ and $t$. We define 
\begin{eqnarray}
y \equiv {a\over a_0} \,, \qquad \tau \equiv H_0(t-t_0) \,.
\end{eqnarray}
Using these variables, we can rewrite Eq.\ (\ref{evoleqn}) in the
following form:
\begin{eqnarray}
\left( {dy \over d\tau} \right)^2 &=& 
{y^2 \over H_0^2} \left[
{8\pi G\over 3}\, {\rho_0 \over y^3} + 
{\Lambda\over 3} - {k\over y^2 a_0^2} \right] \nonumber\\ 
&=& {1\over y} \Omega_m + y^2 \Omega_\Lambda + \Omega_k 
\end{eqnarray}
Eliminating $\Omega_k$ now using Eq.\ (\ref{sum=1}), we obtain
\begin{eqnarray}
\left( {dy \over d\tau} \right)^2 = 1 + \Big( \frac 1y - 1 \Big)
\Omega_m + \Big( y^2-1 \Big) \Omega_\Lambda \,,
\end{eqnarray}
or
\begin{eqnarray}
d\tau = {dy \over \sqrt{1 + \Big( {1\over y} - 1 \Big)
\Omega_m + \Big( y^2-1 \Big) \Omega_\Lambda}} \,.
\label{yeqn}
\end{eqnarray}
If there was a big bang, $y$ was zero at the time of the bang, i.e.,
at $t=0$. On the other hand, $y=1$ now, by definition.  Integrating
Eq.\ (\ref{yeqn}) between these two limits, we obtain
\begin{eqnarray}
H_0t_0 &=& \int_0^1 {dy \over \sqrt{1 + \Big( {1\over y} - 1 \Big)
\Omega_m + \Big( y^2-1 \Big) \Omega_\Lambda}} \,.
\label{yage}
\end{eqnarray}
This is the equation which shows that the age of the universe is {\em
not}\/ independent, but rather is determined by $H_0$, $\Omega_m$ and
$\Omega_\Lambda$. 

\begin{figure}
\centerline{\epsfysize=.3\textheight
\epsfbox{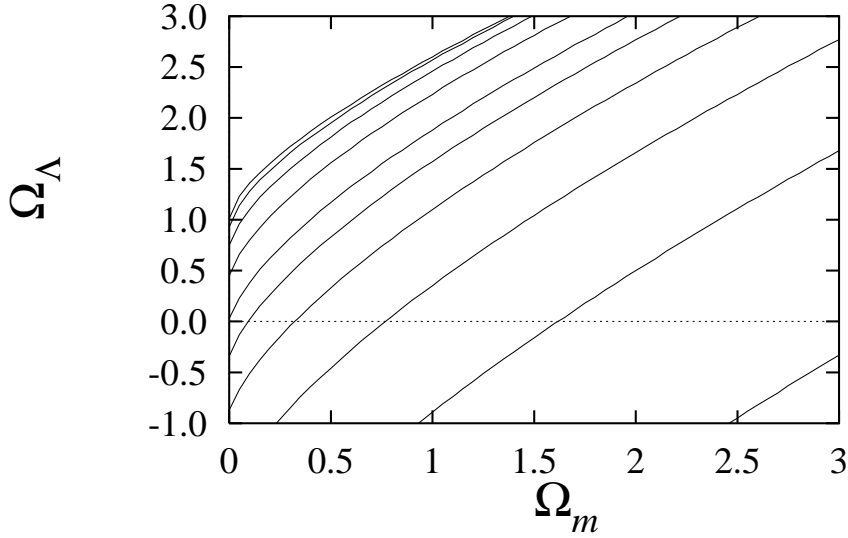}} 
\caption[]{Age plots. The lines shown are contours of equal $H_0t_0$
in the $\Omega_m$-$\Omega_\Lambda$ plane. Starting from the lower
right corner of the plot, the value of $H_0t_0$ for 
different lines are $0.5$, $0.6$, $0.7$, $0.8$, $0.9$, $1.0$, $1.2$,
$1.5$, $2.0$ and $5.0$. The curve for $H_0t_0=\infty$ would be
virtually indistinguishable from that of $H_0t_0=5$ in this
plot.}\label{f:age} 
\end{figure}
Conventionally, one does not use the dimensionless parameter $y$, but
rather uses the {\em red-shift parameter}\/ $z$, defined by
\begin{eqnarray}
1+z \equiv {a_0\over a} = {1\over y} \,.
\label{z}
\end{eqnarray}
Using this variable, Eq.\ (\ref{yeqn}) becomes
\begin{eqnarray}
d\tau = {dz \over 1+z} \; {1\over \sqrt{(1+z)^2 (1+\Omega_m
z) - z(2+z) \Omega_\Lambda}} \,,
\label{zeqn}
\end{eqnarray}
so that 
Eq.\ (\ref{yage}) can be written in the following
equivalent form:
\begin{eqnarray}
H_0t_0 
&=& \int_0^\infty {dz \over 1+z} \; {1\over \sqrt{(1+z)^2 (1+\Omega_m
z) - z(2+z) \Omega_\Lambda}} \,.
\end{eqnarray}

In Fig.~\ref{f:age}, we have shown contours of equal values of
$H_0t_0$ for different values of $\Omega_m$ and $\Omega_\Lambda$. For
a fixed value of $\Omega_\Lambda$, the figure shows that $H_0t_0$
decreases for increasing values of $\Omega_m$. This is because with
more matter, the force of gravity is larger, and the initial bang
slows down in less time. On the other hand, for fixed $\Omega_m$, the
age increases with increasing $\Omega_\Lambda$. And finally, note that
the contour lines for large values of $H_0t_0$ appear very closer and
closer, and approaches asymptotically the line for infinite
age. Beyond this line, there is a region in the parameter plane for
which the age integral diverges. This is the upper left part of the
plot. Later we will discuss what sort of evolution does this part
represent.

\section{Evolution of the universe}\label{ev}
\begin{figure}
\centerline{\epsfysize=.25\textheight \epsfxsize=.5\textwidth
\epsfbox{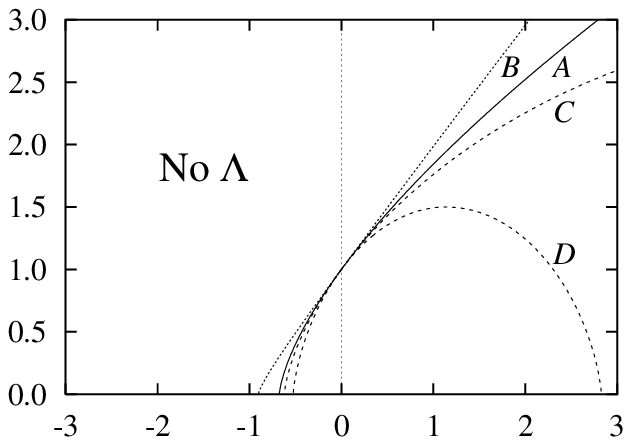}
\epsfysize=.25\textheight \epsfxsize=.5\textwidth
\epsfbox{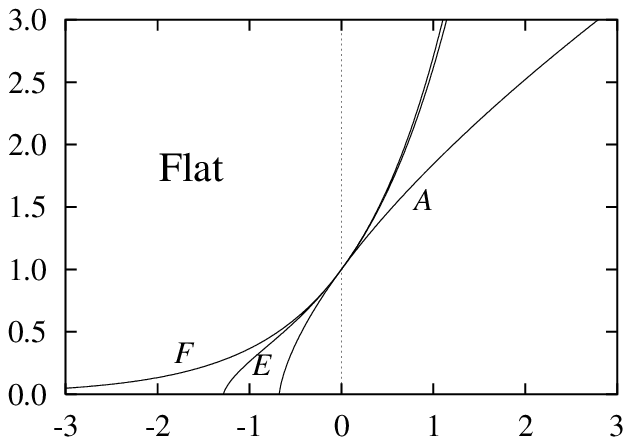}} 
\centerline{\epsfysize=.25\textheight \epsfxsize=.5\textwidth
\epsfbox{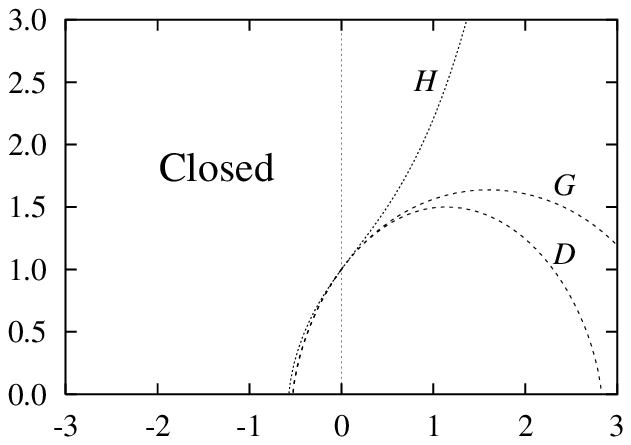}
\epsfysize=.25\textheight \epsfxsize=.5\textwidth
\epsfbox{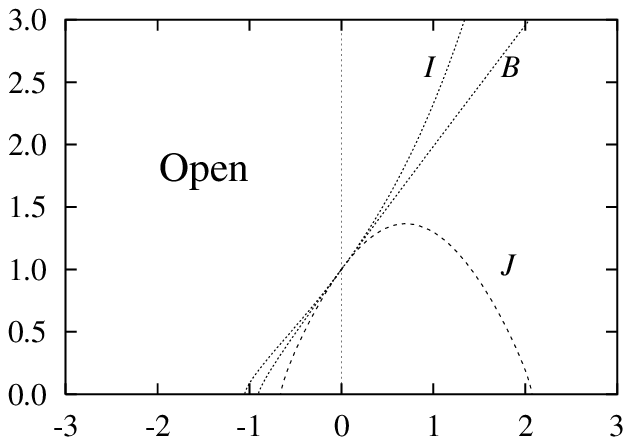}} 
\centerline{\epsfysize=.25\textheight \epsfxsize=.5\textwidth
\epsfbox{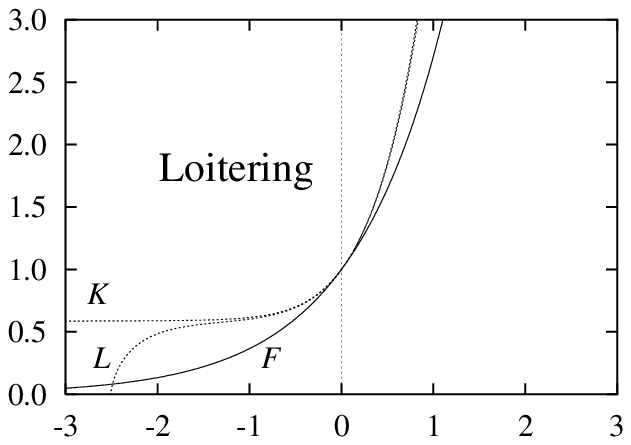}
\epsfysize=.25\textheight \epsfxsize=.5\textwidth
\epsfbox{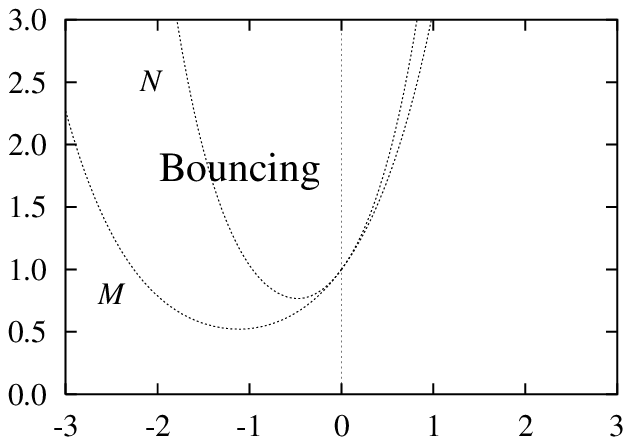}} 
\caption[]{Evolution of the scale parameter with respect to time for
different values of matter density and cosmological parameter. The
horizontal axis represents $H_0(t-t_0)$, the combination which appears
in the evolution equation, Eq.\ (\ref{intyeqn}). The vertical axis is
$y=a/a_0$ in each case. The values of ($\Omega_m$,$\Omega_\Lambda$)
for different plots are: A=(1,0), B=(0.1,0), C=(1.5,0), D=(3,0),
E=(0.1,0.9), F=(0,1), G=(3,.1), H=(3,1), I=(.1,.5), J=(.5,$-1$),
K=(1.1,2.707), L=(1,2.59), M=(0.1,1.5), N=(0.1,2.5).
}\label{f:evol}
\end{figure}

In order to discuss the evolution of the universe, let us not
integrate Eq.\ (\ref{yeqn}) all the way to the initial singularity,
but rather to any arbitrary time $t$. This gives
\begin{eqnarray}
H_0 (t - t_0) 
= \int_0^{y} 
{dy' \over \sqrt{1 + \Big( {1\over y'} - 1 \Big)
\Omega_m + \Big( y'^2-1 \Big) \Omega_\Lambda}} \,.
\label{intyeqn}
\end{eqnarray}
Equivalently, using the red-shift variable, we can write
\begin{eqnarray}
H_0 (t_0 - t) 
&=& \int_0^{z} {dz' \over 1+z'} \; {1\over \sqrt{(1+z')^2 (1+\Omega_m
z') - z'(2+z') \Omega_\Lambda}} \,.
\label{lookback}
\end{eqnarray}
The numerical results of this integration has been shown in
Fig.~\ref{f:evol} for different values of the pair
$(\Omega_m,\Omega_\Lambda)$. 

Some discussion of these evolution plots are in order. We have shown
the plots in six different panels, with some overlaps between
them. The first panel represents different examples of universes with
$\Omega_\Lambda=0$, i.e., no cosmological constant. We know that in
this case, the universe closes and collapses to zero volume in the
future if $\Omega_m>1$. This shows clearly in the curve D, for which
$\Omega_m=3$. For curve C, the collapsing is not clear in the diagram,
but that's because we have not plotted for large enough values of the
time variable. For curve B, $\Omega_m=0.1$, and this universe expands
forever. Curve A has $\Omega_m=1$. Since $\Omega_\Lambda=0$, this
implies that $\Omega_k=0$, i.e., the universe is flat.

In the second panel, we show more flat universes, i.e., universes for
which $k=0$, or alternatively $\Omega_m+\Omega_\Lambda=1$. We see
that, subject to this condition, the larger the value of
$\Omega_\Lambda$, the longer is the past history of the universe. The
extreme example is obtained for $\Omega_\Lambda=1$, shown as curve
F. Although this means $\Omega_m=0$ which is unrealistic, since we
know there must be some matter in the universe, otherwise who is
writing this article and who is going to read it? But in any case,
this choice is instructive, and it shows that in this case, the scale
parameter started in the infinite past with a zero value, and expanded
very slowly for most of the past history of the universe, until
recently when it started to grow.

In the third panel, we show universes with $k>0$, i.e.,
$\Omega_m+\Omega_\Lambda>1$. One of these curves, viz.\ D, has been
encountered in the first panel already. This one has $\Omega_m=3$ and
$\Omega_\Lambda=0$, and it is destined for a recollapse in the
future. If we increases $\Omega_\Lambda$ a little bit, viz.\ to 0.1,
we still obtain a recollapsing universe, as shown by the curve
G. However, if we take a much larger value of $\Omega_\Lambda$, we
find a evolution curve like that shown as H. Here, the universe
expands forever, despite the fact that $k>0$, or $\Omega_k<0$. This is
contrary to our intuition obtained from the case of $\Omega_\Lambda=0$
universes.

Similar conflict is encountered in the fourth panel, which shows open
universes, i.e., universes with $\Omega_m+\Omega_\Lambda<1$. The curve
B, shown earlier in the first panel, shows an evergrowing universe,
and so does curve I. But curve J shows a recollapsing universe.

In view of these, let us adopt a more general classification of the
different kinds of universes that we have seen. We will keep calling a
universe {\em open}\/ or {\em closed}\/ depending on whether
$\Omega_k$ is positive or negative. On the other hand, a universe will
be called {\em elliptic}\/ if it recollapses in the future, and
{\em hyperbolic}\/ if it is evergrowing. For $\Omega_\Lambda=0$, open
universes are necessarily hyperbolic and closed universes are
elliptic. But for non-zero $\Omega_\Lambda$, we can have open
hyperbolic, open elliptic, closed hyperbolic or closed elliptic
universes.

Before we move on to the next section, we need to discuss the two
lower panels of Fig.~\ref{f:evol}. In the lower left panel, we have
repeated curve F which was shown in the second panel of flat
universes. As we said, in this case the universe spends an infinite
amount of time for which it hardly expands. Curve K gives another
example of this kind. For different values of $\Omega_m$ and
$\Omega_\Lambda$, the initial scale parameter is different. These are
examples of {\em loitering}\/ universes. Curve L shows a universe
which is very close to the loitering case. It stays at a particular
value of the scale parameter for a long time, but not for infinite
amount of time.

If the value of $\Omega_\Lambda$ is increased further from those of
any of the loitering universe cases in the previous panel, we find
{\em bouncing}\/ universes, as shown in the last panel. In these
cases, the universe started with an infinite scale factor, shrunk to a
minimum value at some time in the past and is expanding now. This
expansion will go on forever. These are the cases for which the age
integral in Eq.\ (\ref{yage}) diverges, and is represented by the
empty upper left corner of the age plots of Fig.~\ref{f:age}.

\begin{figure}
\centerline{\epsfysize=.35\textheight
\epsfbox{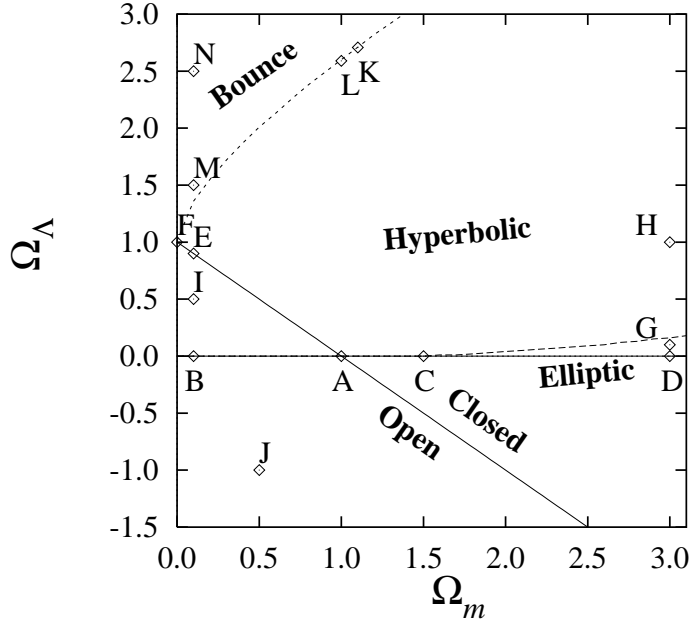}} 
\caption[]{Summary of different evolution plots presented in
Fig.~\ref{f:evol}. The borderline between open and closed universes
represents flat universes, between hyperbolic and
bounce universes represents loitering universes, and the one between
elliptic and hyperbolic universes are called critical
universes.}\label{f:scheme} 
\end{figure}
All these panels have been summarized in Fig.~\ref{f:scheme}, where we
have shown which values of $\Omega_m$ and $\Omega_\Lambda$ give open
or closed universes, as well as hyperbolic or elliptical universes. As
we see clearly in this plot, the identification of closed universes
with elliptical ones works only for $\Lambda=0$, and so does the
identification of open universes with hyperbolic ones.

\section{Strategies}\label{st}
We have seen, with many examples, the nature of the evolution of the
universe for various values of the cosmological parameters. Now the big
question is: how to determine which one is our own universe?

One thing is easy to say, that we do not live in a bouncing
universe. The reason is that these universes have a minimum value of
$a/a_0$, i.e., a certain maximum value for the red-shift parameter
$z$, say $z_{\rm max}$. This value satisfies the relation
\begin{eqnarray}
z_{\rm max}^2 \big( z_{\rm max} + 3) \leq {2\over \Omega_m} \,.
\end{eqnarray}
We know there exists quasars at $z>4$. So, if we make the reasonable
estimate that $\Omega_m>0.02$, these cosmologies are ruled
out. However, this still leaves us with a vast region in the
$\Omega_m$-$\Omega_\Lambda$ parameter region, and we need to set up a
strategy to proceed.

The general (and obvious) strategy is the following. We need to find
some observable quantity which depends on the cosmological parameters,
determine the value of that observable. That will give us the values
for the cosmological parameters.  Our original question now shifts to
the following one: what is a good observable for this purpose?

We have already encountered one quantity which depends on the
cosmological parameters, viz., the age of the universe. Unfortunately,
it cannot serve our needs very well. The reason is that one cannot
really determine $t_0$ from any direct observations. No matter how old
an object is found in the universe, that will not determine $t_0$, but
rather only put a lower limit on it. So we will use estimates of $t_0$
only as a check for whatever strategy we adopt.

Another quantity that depends on $\Omega_m$ and $\Omega_\Lambda$ is
the {\em lookback time}\/ for any object, which is $t_0-t$ that
appears in, e.g., Eq.\ (\ref{lookback}). We look at a certain object
in the sky. We find its red-shift, which can be done very accurately.
If we now have some way of knowing the time at which the light that we
observe was emitted from the object, we are through. But this cannot
be done very well. By looking at the composition of the object and by
comparing it with some evolution model, we can probably estimate the
age, but it will depend on the evolution model, and so the method is
somewhat uncertain.

Therefore one usually relies on the measurement of distances of
objects as a function of their redshifts for the purpose of
determining cosmological constants. In order to explain the process,
first we need to find out how the physical distance of an object
depends on $z$ for different values of $\Omega_m$ and
$\Omega_\Lambda$. This issue is taken up in the next
section. Following that, in Sec~\ref{dm}, we will discuss some methods
of measurements of distances, and summarize how they determine the
cosmological constants. The importance of these determinations for
particle physics is given in Sec.~\ref{im}.

\section{Distance vs redshift relation}\label{dz}
A light ray traces a null geodesic, i.e., a path for which $ds^2=0$ in
Eq.\ (\ref{FRWmetric}). Thus, a light ray coming to us satisfies the
equation 
\begin{eqnarray}
{dr\over dt} = {\sqrt{1-kr^2} \over a} \,,
\end{eqnarray}
where $r$ is the dimensionless co-ordinate distance introduced in Eq.\
(\ref{FRWmetric}). Using Eqs.\ (\ref{z}) and (\ref{zeqn}), we can
rewrite it as
\begin{eqnarray}
{dr\over \sqrt{1 + \Omega_k a_0^2 H_0^2 r^2}} = (1+z) {dt \over a_0} 
= {1\over H_0a_0} \; {dz\over \sqrt{(1+z)^2 (1+\Omega_m
z) - z(2+z) \Omega_\Lambda}} \,,
\end{eqnarray}
where on the left side, we have replaced $k$ by $\Omega_k$ using the
definition of Eq.\ (\ref{Omegas}).  Integration of this equation
determines the co-ordinate distance as a function of $z$:
\begin{eqnarray}
H_0 a_0 r(z) = {1\over \sqrt{|\Omega_k|}} \; {\rm sinn} \Bigg[
\sqrt{|\Omega_k|} \int_0^z {dz'\over \sqrt{(1+z')^2 (1+\Omega_m
z') - z'(2+z') \Omega_\Lambda}} \Bigg] \,,
\end{eqnarray}
where ``sinn'' means the hyperbolic sine function if $\Omega_k>0$, and
the sine function if $\Omega_k<0$. If $\Omega_k=0$, the sinn and the
$\Omega_k$'s disappear from the expression and we are left only with
the integral.

\begin{figure}
\centerline{\epsfysize=.3\textheight
\epsfbox{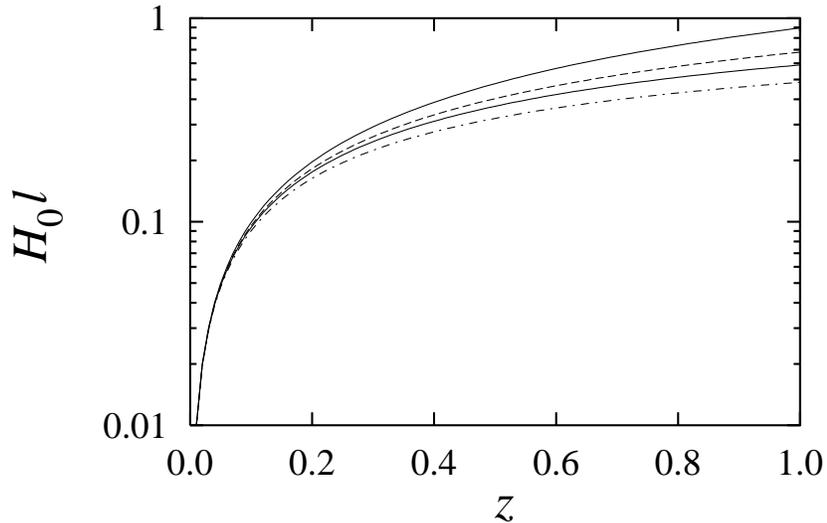}} 
\caption[]{$H_0\ell$ vs.\ red-shift plots, where $\ell$ is the
luminosity distance. From bottom to top, the plots correspond to the
choices of $\Omega_m$ and $\Omega_\Lambda$ denoted by D, A, B, and E
in the caption of Fig.~\ref{f:evol}.  }\label{f:distance}
\end{figure}
The physical distance to a certain object can be defined in various
ways.\footnote{See, e.g., Sec.~3.3 of Ref.~\cite{CPT}, or Sec.~19.9 of
Ref.~\cite{lightman}.} For what follows, we will need what is called
the ``luminosity distance'' $\ell$, which is defined in a way that the
apparent luminosity of any object goes like $1/\ell^2$. This is
related to the co-ordinate distance \cite{lightman} by
\begin{eqnarray}
\ell(z) = a_0^2 r(z) / a(z) = (1+z) a_0 r \,.
\end{eqnarray}
Thus,
\begin{eqnarray}
H_0 \ell(z) = {1+z\over \sqrt{|\Omega_k|}} \; {\rm sinn} \Bigg[
\sqrt{|\Omega_k|} \int_0^z {dz'\over \sqrt{(1+z')^2 (1+\Omega_m
z') - z'(2+z') \Omega_\Lambda}} \Bigg] \,.
\end{eqnarray}
In Fig.~\ref{f:distance}, we plot $H_0\ell(z)$ for various choices of
$\Omega_m$ and $\Omega_\Lambda$ to show the general nature of the
variation.

\section{Measurement of distances}\label{dm}
\subsection{General remarks}
So let us now ask how one can measure the distance of an object. In a
sense, this is the most important problem of observational cosmology.
Older methods of measuring distances used ladder techniques
\cite{bothun}. This means that, upto a certain distance, one
particular method was used, and this was used to calibrate another
method which could be used for larger distances. This process was
carried on to higher and higher rungs of the ladder. In order to get
to distances large enough to distinguish between different universes,
one had to go through several rungs of the ladder, and accordingly
there were too many uncertainties in the measurement.

Recently various other methods have been developed and employed to
measure the cosmological parameters. These methods avoid using the
ladder technique and are therefore expected to provide much better
estimates of distances. Some of these are listed below.
\begin{description}

\item {\bf Gravitational lensing}. Different images of an object from
a gravitational lens are formed by light rays coming through different
directions, and therefore traversing different path lengths. If the
original object has some periodicity in the luminosity, the
periodicity of the images will depend on the path length. This
provides a way of determining distances \cite{gravlens}. So far, data
is very scarce.

\item {\bf Sunayev-Zeldovich effect}. Photons from the cosmic
microwave background, when passing through a galaxy, get
scattered. As a result, they gain in energy. Thus, looking at the
direction of a galaxy, the microwave background radiation does not
look thermal. Rather, there is some depletion of the low energy
photons and some gain of the high energy ones. The amount of this
depletion depends on the size of the galaxy. Using this, one can
measure the actual sizes of the galaxies. The apparent size observed
would then provide a measure for the distance \cite{SZeffect}. The
data obtained until now show a lot of scatter, and we will not discuss
them here.

\item {\bf Microwave background anisotropies}. The evolution of
density perturbations depend on the values of the cosmological
parameters. Therefore, the anisotropies in the microwave background
radiation should be predictable in terms of the values of $\Omega_m$
and $\Omega_\Lambda$. With very accurate determinations of these
anisotropies in the near future, this method will probably turn out to
be the best probe for the cosmological parameters.

\item {\bf Supernova 1a.} In this method, it is assumed that any
supernova of type 1a has the same absolute luminosity. The measurement
of the apparent luminosity therefore provides a measure for the
distance. This will be discussed in more detail in the remaining part.

\end{description}

\begin{figure}
\centerline{\epsfxsize=.5\textwidth 
\epsfysize=.25\textheight
\epsfbox{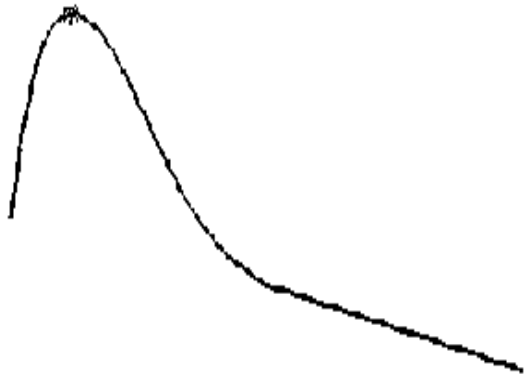}} 

\caption[]{Schematic lightcurve for a supernova of type
1a.}\label{f:lcurve}
\end{figure}
\subsection{Results from Supernova 1a measurements}\label{sn1a}
Supernova of type 1a are identified by the nature of their light
curve, i.e., the variation of their intensity with time, as shown in
Fig.~\ref{f:lcurve}. In addition, the spectral analysis reveals lines
for heavy elements like magnesium and silicon.

Supernovas of this type are believed to occur due of the merger of
two white dwarf stars whose masses are very close to the Chandrasekhar
limit. Since the masses entering into the explosion is roughly the
same for any supernova 1a, it is reasonable to assume that the
intrinsic (or absolute) luminosity of any supernova of this type is
the same. Thus, if one measures the apparent luminosity at the maximum
of the light curve, this should scale as the inverse square of the
distance of the supernova, where the distance is defined to be the
luminosity distance. Astronomers use `magnitudes' to denote
luminosities, which is a logarithmic representation. Thus, the
apparent magnitude at the maximum should satisfy the relation
\begin{eqnarray}
m_{\rm max} \propto \log \ell^2 \,.
\end{eqnarray}
We have already shown in Fig.~\ref{f:distance} how $\log \ell$ varies
with $z$ for different values of $\Omega_m$ and $\Omega_\Lambda$. So
the strategy must be simple. One should plot the observed values of
the distance vs $z$, and determine which pairs of values of $\Omega_m$
and $\Omega_\Lambda$ give the best fit for the observed points.

There is only one small point to settle before embarking on this
program. The vertical axis of Fig.~\ref{f:distance} gives the
logarithmic values of $H_0\ell$, not of $\ell$. Thus, unless we know
what $H_0$ is, we cannot proceed.  However, the thing to note is that
for small values of $z$, the plots are independent of $\Omega_m$ and
$\Omega_\Lambda$. Thus, if one determines the red-shifts and apparent
magnitudes of nearby supernovas, that should determine the Hubble
parameter. Using this value then, one can go over to larger values of
$z$ and determine $\Omega_m$ and $\Omega_\Lambda$.

\begin{figure}
\centerline{\epsfysize=.4\textheight 
\epsfbox{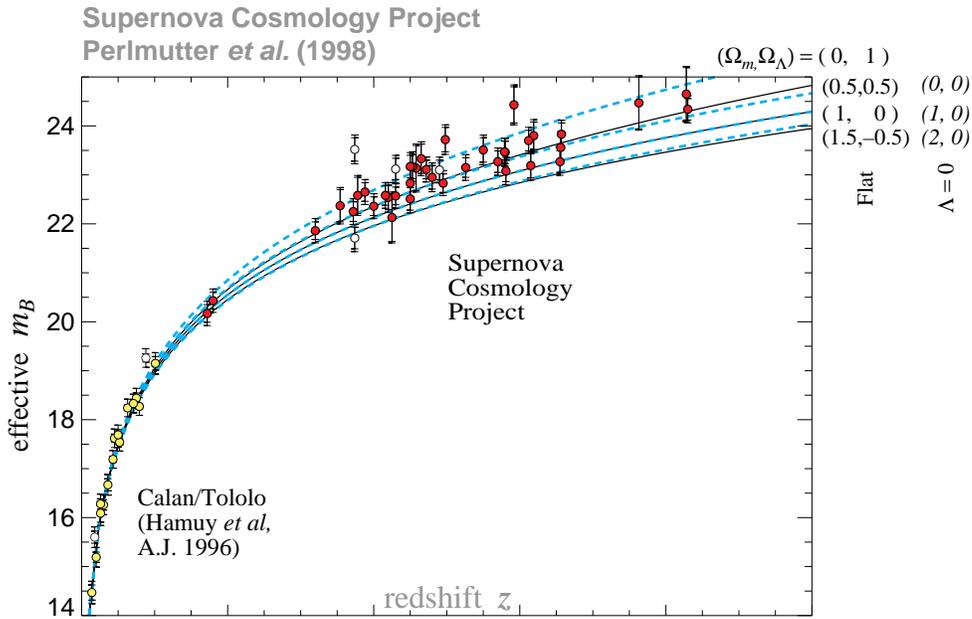}} 
\caption[]{Results of redshift vs. apparent luminosity of a number of
type-1 supernovas. The horizontal scale goes from 0 to 1. The data for
low redshift are from Ref.~\cite{CalanTololo}, and for $z>0.2$ is from
Ref.~\cite{SCP}. The figure is taken from Ref.~\cite{SCP}, with slight
modification of the legends. The dashed lines are for flat universes,
and the solid lines for $\Lambda=0$ universes, with values of
$\Omega_m$ and $\Omega_\Lambda$ given at the top right corner.
}\label{f:lblplot}
\end{figure}
\begin{figure}
\centerline{
\epsfysize=.4\textheight
\epsfbox{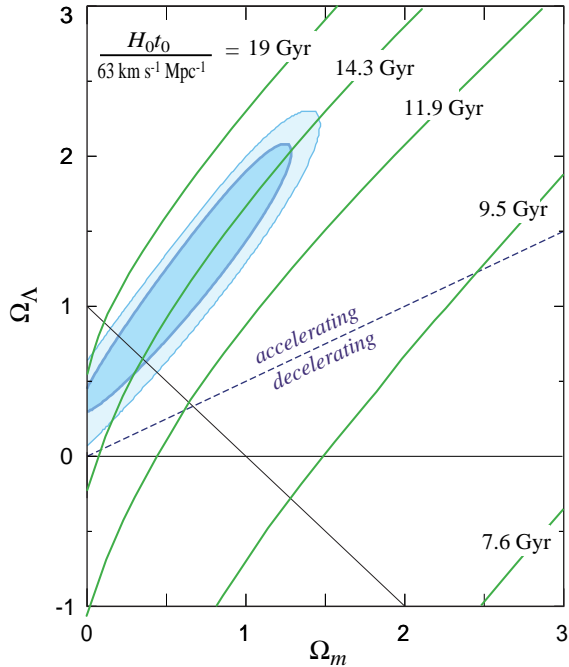}} 
\caption[]{Best fits for the data of Fig.~\ref{f:lblplot} in the
$\Omega_m$-$\Omega_\Lambda$ plane. Superimposed are the contours of
equal age, with the value of the Hubble parameter set at $63\
\rm km\, s^{-1} \, Mpc^{-1}$.}\label{f:lblresults}
\end{figure}

The part of the task at low $z$ was extensively done a few years ago
\cite{CalanTololo}, and the data are shown in
Fig.~\ref{f:lblplot}. The value of the Hubble parameter obtained from
this data is:
\begin{eqnarray}
H_0 = (63.1 \pm 3.4 \pm 2.9) \ \rm km\, s^{-1} \, Mpc^{-1} \,.
\label{H0result}
\end{eqnarray}
Using this value, one can determine the distances of farther
supernovas. This has been started in 1998 under the team effort called
the `Supernova Cosmology Project' (SCP). They measured the redshift
and the effective magnitude of 42 supernovas and published their
result, which is shown in Fig.~\ref{f:lblplot}. Superimposed on their
data are the results expected from different combinations of
$\Omega_m$ and $\Omega_\Lambda$. Apart from an overall normalization,
these are the curves as shown in Fig.~\ref{f:distance}. One can now
determine which values of $\Omega_m$ and $\Omega_\Lambda$ fit the data
sufficiently well, and the results of the analysis of the SCP has been
shown in Fig.\ref{f:lblresults}.

Clearly, the fits show that the cosmological constant is non-zero. In
fact, if we stick to flat universes, which are denoted by the diagonal
straight line on this plot, the best value that the SCP advocate is
\begin{eqnarray}
\Omega_m \approx 0.22 \,, \qquad \Omega_\Lambda \approx 0.78 \,.
\label{best}
\end{eqnarray}
In general, however, we obtain a region in the plane, and the regions
allowed at different confidence levels are shown on the plot.

In Fig.~\ref{f:age}, we have seen how these contours look like. In
general, one obtains contours of equal $H_0t_0$. Assuming a value of
$H_0$, one can draw contours of equal age, which have also been
superposed on the plot of Fig.~\ref{f:lblresults}. The value of $H_0$
assumed is inspired by the results shown in Eq.\ (\ref{H0result}), and
is indicated in the figure and the caption. The best fit that the SCP
advocate is:
\begin{eqnarray}
t_0 = (14.9 \pm 1) \times 10^9 \rm yr \,,
\label{t0flat}
\end{eqnarray}
for flat universes. If one does not assume a flat universe, the best
fit for the age is
\begin{eqnarray}
t_0 = (14.5 \pm 1) \times 10^9 \rm yr \,.
\label{t0general}
\end{eqnarray}
Of course, both these values for $t_0$ are for the value of $H_0$
indicated in the figure, which is the central value quoted in Eq.\
(\ref{H0result}).

\section{Implications}\label{im}
What do these result mean for particle physics? Various particles,
like neutrinos, can contribute only to $\Omega_m$. The central value
of the Hubble parameter, taken literally, tells us that the critical
density of the universe is
\begin{eqnarray}
\rho_c = 4 \; {\rm keV/cm}^3 \,.
\end{eqnarray}
Moreover, taking the best values advocated in Eq.\ (\ref{best}), we
obtain that the matter density in the present universe is
\begin{eqnarray}
\rho_0 \approx 900 \; {\rm eV/cm}^3 \,.
\end{eqnarray}
With the standard number density of neutrinos of about 110/cm$^3$, the
maximum allowed mass for light stable neutrinos is about 8\,eV. One
can similarly go through various other constraints on neutrino
properties derived from cosmology, and find that they become much more
stringent than the values usually quoted.

And this is not all. As we already pointed out, the growth of density
perturbations have different characteristics in a universe with
non-zero $\Lambda$. These considerations also put extra constraints on
different types of possible dark matter. This is a subject which is
only beginning to be investigated~\cite{Primack}.




\begin{thebibliography}{WW}
\bibitem{CPT} S.~M. Carroll, W.~H. Press, E.~L. Turner:
Ann. Rev. Astron. Astrophys. 30 (1992) 499.

\bibitem{lightman} A.~P. Lightman, W.~H. Press, R.~H. Price,
S.~A. Teukolsky: {\sl Problem book in relativity and gravitation},
(Princeton University Press, 1975).

\bibitem{bothun} For a discussion of these older techniques, see,
e.g., G. Bothun: {\sl Modern cosmological observations and problems},
(Taylor \& Francis, 1998).

\bibitem{gravlens} For details, see, e.g., R.~D. Blandford and
R. Narayan: Ann. Rev. Astron. Astrophys. 30 (1992) 311.

\bibitem{SZeffect} For details, see, e.g., Y. Rephaeli:
Ann. Rev. Astron. Astrophys. 33 (1995) 541; M. Birkinshaw:
Phys. Rep. 310 (1999) 97.

\bibitem{CalanTololo} M. Hamuy, M.~M. Phillips, N.~B. Suntzeff,
R.~A. Schommer, J. Maza, R. Avil\'es: Astronom. J. 112 (1996) 2391.

\bibitem{SCP} The Supernova Cosmology Project, S. Perlmutter et al,
{\tt astro-ph/9812133}.

\bibitem{Primack} See, e.g., J.R.~Primack and M.A.~Gross,
astro-ph/9810204.

\end{thebibliography}
\end{document}